\documentclass[a4paper,11pt]{article}
\usepackage{jheppub}
\usepackage{eurosym}
\usepackage{amssymb}
\usepackage{amsfonts,cite}
\usepackage{amsmath}
\usepackage{multirow}
\usepackage{epsfig}
\usepackage[utf8]{inputenc}
\usepackage{multicol}
\usepackage{graphicx}
\usepackage{fancyhdr}
\usepackage{placeins}
\usepackage{slashed}
\usepackage{lipsum}
\usepackage{array}
\usepackage{amscd}
\usepackage{tikz-cd}
\usepackage[thinlines]{easytable}
\usepackage{mathrsfs}
\usepackage[export]{adjustbox}
\usepackage{blindtext}
\usepackage{wrapfig}

\newbox\mybox

\newcommand\fverb{\setbox\mybox=\hbox\bgroup\verb}
\newcommand\fverbdo{\egroup\medskip\noindent\fbox{\unhbox\mybox}\ }
\newcommand\fverbit{\egroup\item[\fbox{\unhbox\mybox}]}
\newcommand{\dd}{\mathrm{d}}

\abstract{We investigate nonlinear excitations of the discrete shape mode about a static complex soliton of the integrable Bullough-Dodd model. Extending the linear stability analysis to third order in the excitation amplitude, we construct explicit spatially localised corrections to the shape-mode oscillation and determine its nonlinear frequency shift using the Poincar\'e-Lindstedt method. Although both the second and third harmonics lie inside the continuous spectrum of the linearised fluctuation operator, the corresponding inhomogeneous equations admit exact localised particular solutions. We demonstrate that the nonlinear sources have vanishing on-shell overlaps with the scattering states at both harmonic frequencies, suppressing the corresponding radiative responses. We further investigate the role of integrability and show how a higher commuting symmetry of the Bullough-Dodd equation provides a perturbative selection-rule interpretation of these cancellations. Our results illustrate how higher integrable symmetries can constrain nonlinear resonances that are otherwise permitted by the linear spectrum.}

\title{Higher-symmetry selection rules for second- and third-harmonic radiation in the Bullough-Dodd model}

\author{Andreas Fring}

\affiliation{Department of Mathematics, City St George's, University of London,  Northampton Square, \\ London EC1V 0HB, UK}

\emailAdd{a.fring@city.ac.uk}	


\begin{document}
	\maketitle
	
\pagestyle{fancy}
\fancyhead{} 
\fancyhead[LE,RO]{\small\itshape  Higher-symmetry selection rules for second- and third-harmonic radiation in the BD model} 
	
\renewcommand{\headrulewidth}{0.4pt}

\section{Introduction}

The stability of soliton solutions is a central question in nonlinear field theory. While linear stability analysis determines the spectrum of small fluctuations about a given background, it does not necessarily provide a complete description of the nonlinear dynamics. In particular, a soliton may possess discrete internal oscillations, commonly referred to as shape modes, whose nonlinear interactions generate higher harmonics. When the frequencies of these harmonics lie inside the continuous spectrum of the linearised fluctuation operator, they can excite propagating waves and transfer energy from the localised excitation into the continuum. This mechanism provides a possible source of radiative damping of internal oscillations.

The nonlinear coupling of internal modes to continuum radiation has been investigated extensively for the $\phi^4$ kink. In particular, Manton and Merabet \cite{manton1997kinks} calculated the radiation generated at twice the shape-mode frequency and examined the resulting decay of the internal oscillation. Subsequent studies provided a systematic perturbative description of this radiative damping \cite{barashenkov2009wobbling}. The role of vibrational modes in kink-antikink interactions has also been investigated using collective-coordinate approximations \cite{takyi2016coll}, while more recent work has explicitly incorporated continuum radiation modes into this framework \cite{navarro2023inclusion}. Rigorous results concerning the asymptotic stability of kinks have further highlighted the importance of nonlinear resonances and the coupling between discrete and continuous modes \cite{kowalczyk2017kink,KowalczykMartel2024}. More generally, such energy transfer can be understood through nonlinear resonance theory and the Fermi golden rule mechanism \cite{soffer1999resonances}. A central quantity in these analyses is the overlap of the nonlinear forcing with the scattering states at the resonant frequency, which determines the coupling to the corresponding radiation channels. While a non-vanishing overlap generally allows radiation to be emitted, its exact cancellation can suppress the radiative response at the relevant perturbative order.

The situation is particularly interesting in integrable field theories. Integrability supplies additional conserved quantities and commuting flows, which impose constraints on the nonlinear dynamics beyond those following from energy conservation alone. The role of higher conserved charges in restricting inelastic scattering processes was demonstrated by Parke \cite{Parke1980}, while more recent work has investigated the cancellation of particle-production amplitudes in affine Toda field theories \cite{dorey2022tree}. Although these results concern scattering processes rather than radiation from soliton shape modes, they illustrate how integrability can suppress otherwise permitted nonlinear interactions. Another manifestation of integrability is the existence of exact localised time-periodic solutions, such as the breathers of the sine-Gordon equation, whose integrable structure can be analysed by inverse scattering \cite{AKNSsg}. It is therefore natural to ask whether the additional symmetries of an integrable field theory can also suppress the radiative coupling of discrete shape modes to continuum excitations.

This raises an important distinction between integrable and nonintegrable soliton dynamics. For the nonintegrable $\phi^4$ kink, the internal shape mode couples to continuum radiation at its second harmonic, leading to radiative damping \cite{manton1997kinks,barashenkov2009wobbling}. Integrability may constrain such couplings, but their cancellation does not follow merely from the existence of conserved quantities. In particular, when a soliton possesses an internal mode whose higher harmonics lie inside the continuum, it remains necessary to establish whether the corresponding nonlinear radiation amplitudes vanish.

In this work we address this question for the Bullough-Dodd model, an integrable nonlinear Klein-Gordon type theory with an exponential interaction. The model possesses complex soliton solutions whose stability properties have been investigated in \cite{CorreaFringTaira}. In particular, the linearised fluctuation operator around the static type-I soliton admits a discrete shape mode with frequency $\omega_1=3/2$, while its continuous spectrum is given by $E=3+\kappa^2$. Consequently, the second and third harmonics of the shape mode both lie inside the continuous spectrum. From the viewpoint of linear spectral theory, radiative channels are therefore available at both frequencies.

\begin{wrapfigure}{r}{0.60\textwidth}
	\centering
	\includegraphics[width=\linewidth]{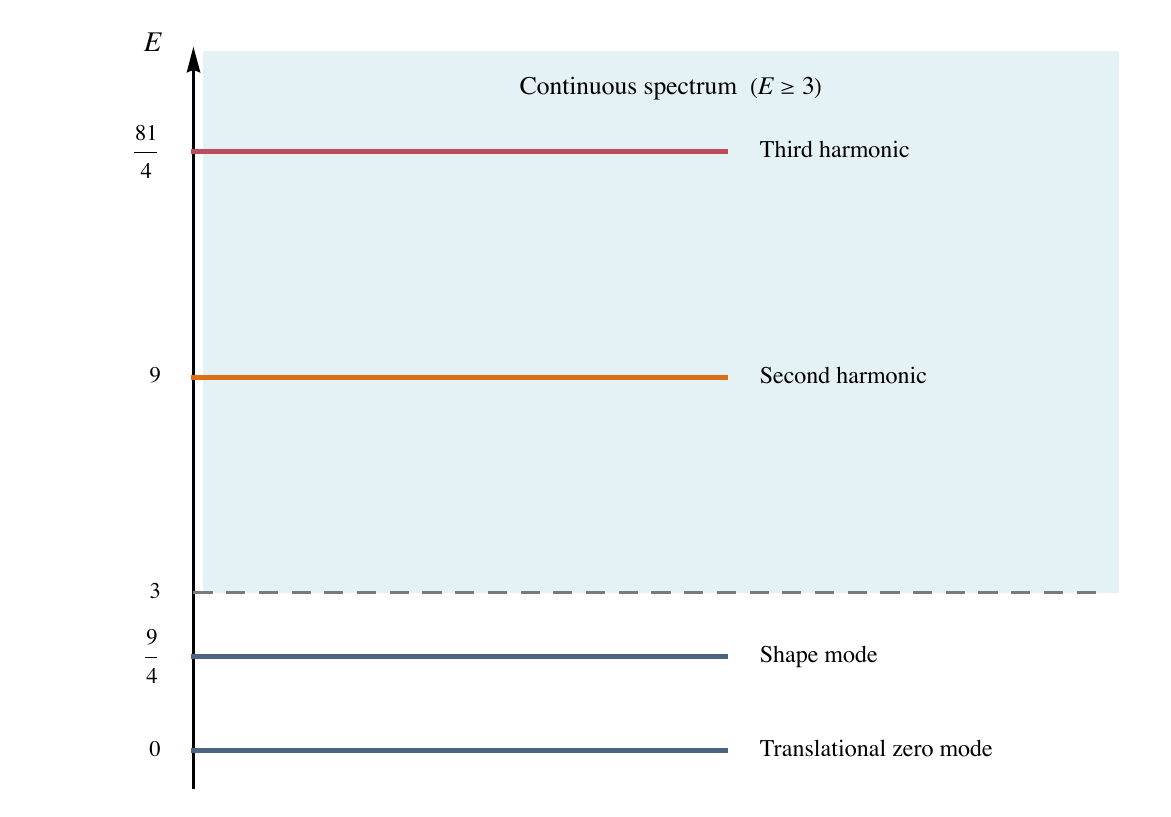}
	\caption{Schematic spectrum of the linearised fluctuation operator. The shaded region denotes the continuum, while the coloured lines indicate the second- and third-harmonic forcing energies. Not to scale.}
	\label{levels}
\end{wrapfigure}

As illustrated in figure~\ref{levels}, the second and third harmonics of the shape mode correspond to $E_2=4\omega_1^2=9$ and $E_3=9\omega_1^2=81/4$, respectively. Both lie strictly inside the continuous spectrum, so propagating radiation channels are available at these frequencies. The harmonic energies represent nonlinear forcing frequencies rather than additional discrete eigenvalues.

We investigate the nonlinear excitation of this shape mode perturbatively up to third order. At second order, we obtain explicit localised solutions for the static correction and the second harmonic. At third order, the nonlinear forcing also contains a component at the fundamental frequency, requiring a Poincar\'e-Lindstedt expansion to eliminate secular contributions. The resulting frequency correction and the third-harmonic spatial profile can again be determined analytically. Remarkably, despite their frequencies lying inside the scattering continuum, both higher harmonics admit exponentially localised particular solutions.

We demonstrate that the nonlinear forcing terms have exactly vanishing on-shell overlaps with the corresponding scattering eigenfunctions. These cancellations follow directly from Lagrange's identity and the existence of the localised particular solutions. We then investigate their origin in the integrable hierarchy of the Bullough-Dodd equation. Using a fifth-order commuting symmetry, we obtain perturbative selection-rule constraints arising from the mismatch between the higher-flow eigenvalues of the shape mode and the resonant continuum states. This provides a structural interpretation of the suppression of the second- and third-harmonic couplings, subject to the appropriate interpretation of the continuous spectral projections.

Our analysis concerns particular solutions of the perturbative equations and their coupling to the continuum. It does not establish the existence of an exact nonlinear breather or nonlinear stability against arbitrary perturbations. Rather, it identifies a mechanism through which the integrable structure constrains nonlinear radiation in the vicinity of a complex soliton background.

The paper is organised as follows: In section 2, we introduce the complex Bullough-Dodd soliton and develop the perturbative shape-mode expansion up to and including third order. In section 3, we establish the vanishing of the on-shell radiation amplitudes at the second and third harmonics. Section 4 investigates the higher commuting symmetries of the Bullough-Dodd hierarchy and their implications for nonlinear mode coupling. Our conclusions are presented in section 5.  An appendix establishes a higher conservation law and its algebraic connection to the fifth-order generalised symmetry.

\section{Perturbative excitation of the shape mode}
	
We consider the Bullough-Dodd equation, also known as the Tzitzeica equation, which originally arose in the differential geometry of surfaces and was subsequently investigated as an integrable nonlinear field theory by Dodd and Bullough, and independently by Zhiber and Shabat \cite{tzitzeica1908,BDodd,zhiber1979klein}. The model belongs to the class of integrable affine Toda field theories and is associated with the twisted affine Lie algebra $A_2^{(2)}$. It can be obtained through a $\mathbb{Z}_2$ folding of the simply laced $A_2^{(1)}$ affine Toda theory, within the framework of Dynkin diagram reductions developed in \cite{DIO1983}. Its equation of motion reads
\begin{equation}\label{ BD}
		\phi_{tt}-\phi_{xx}+e^\phi-e^{-2\phi}=0.
\end{equation}
The model is integrable and admits a rich variety of exact solutions. Its integrable structure has been investigated using B\"acklund transformations and higher symmetries \cite{andreevback}, while explicit one- and two-soliton solutions, as well as breathers, have been constructed using a general affine Toda formulation \cite{assis2008bullough}. Its quantum integrable structure has also been explored through the exact form factors of local operators \cite{Fring:1992pj}. 
More recently, complex-valued soliton solutions with real energies have been investigated, including configurations with unbroken and broken $\mathcal{PT}$ symmetry and distinct linear stability properties \cite{CorreaFringTaira}. Their behaviour at second order in perturbation theory has also been examined in \cite{Amila}.

Here we focus on the static positive type-I soliton
\begin{equation}\label{background}
	\phi_s(x)=\log\left(\frac{C+2}{C-1}\right), \qquad \zeta=\beta+\sqrt3\,x, \qquad C=\cosh\zeta.
\end{equation}

The complex shift $\beta$ is chosen such that the soliton and its fluctuation profiles are nonsingular on the real spatial axis. For $\beta=i\pi/2$, which we use throughout, we have $C=i\sinh(\sqrt3\,x)$. The argument $z=(C+2)/(C-1)$ of the logarithm in \eqref{background} winds once anticlockwise around the origin as $x$ runs along the real axis. We therefore choose the logarithmic branch with its cut along the positive real axis, corresponding to $0<\operatorname{Arg}z<2\pi$. This makes $\phi_s(x)$ continuous for all real $x$, approaching the vacuum values $0$ and $2\pi i$ at the two spatial infinities, respectively. These vacua are related by the discrete symmetry $\phi\mapsto\phi+2\pi i$ of the Bullough-Dodd equation.
	
To study small fluctuations, we expand the field as
\begin{equation}\label{generalexpansion}
		\phi(x,t)=\phi_s(x)+\varepsilon\,\delta\phi_1(x,t) +\varepsilon^2\,\delta\phi_2(x,t) +\varepsilon^3\,\delta\phi_3(x,t)+O(\varepsilon^4).
\end{equation}
Substituting into equation~\eqref{ BD} and comparing powers of $\varepsilon$ gives the equations
\begin{eqnarray}
			(\partial_t^2+H)\delta\phi_1&=&0,   \label{pert1}\\
			(\partial_t^2+H)\delta\phi_2 &=&-\frac12 W(x)(\delta\phi_1)^2,  \label{pert12}\\
			(\partial_t^2+H)\delta\phi_3 &=&-W(x)\delta\phi_1\delta\phi_2 -\frac16 Z(x)(\delta\phi_1)^3,    \label{pert3}
\end{eqnarray}
where
\begin{eqnarray}
			H &=&-\partial_x^2+U(x),\\
			U(x)&=&e^{\phi_s}+2e^{-2\phi_s} =\frac{3(C^3+6C+2)}{(C-1)(C+2)^2},\\
			W(x)&=&e^{\phi_s}-4e^{-2\phi_s} =-\frac{3(C^3-6C^2-4)}{(C-1)(C+2)^2},\\
			Z(x)&=&e^{\phi_s}+8e^{-2\phi_s} =\frac{9C(C^2-2C+4)}{(C-1)(C+2)^2}.
\end{eqnarray}
These equations provide the perturbative hierarchy for fluctuations about the static soliton background.
	
\subsection{Linearised fluctuations and the shape mode}
	
The linear stability problem was analysed in \cite{CorreaFringTaira}. Since the background is static, the first-order equation can be separated by writing
\begin{equation}\label{separation}
		\delta\phi_1(x,t)=u(x)T(t).
\end{equation}
Substituting into (\ref{pert1}) leads to the spatial and temporal equations 
\begin{equation}\label{SL}
			Hu(x)=\lambda u(x), \qquad  \quad	\frac{\dd^2T(t)}{\dd t^2}+\lambda T(t)=0,
\end{equation}
where $\lambda$ is the separation constant. The first equation is a Sturm-Liouville type eigenvalue problem for the spatial fluctuation profile. For a positive eigenvalue $\lambda$, the temporal factor is oscillatory,
\begin{equation}\label{timefactor}
		T(t)=c_1\cos(\omega t)+c_2\sin(\omega t), \qquad \lambda=\omega^2.
\end{equation}
Thus the eigenvalues of $H$ determine the squared fluctuation frequencies. Positive real eigenvalues correspond to oscillatory time dependence, whereas negative or non-real eigenvalues can give rise to exponentially growing perturbations.
	
As established in\cite{CorreaFringTaira}, the discrete spectrum contains a translational zero mode and a shape mode, whose eigenfunctions can be chosen as
\begin{equation}\label{modes}
		\begin{aligned}
			\chi_0(x)&=-\frac{3\coth(\zeta/2)}{C+2}, 	&H\chi_0&=0,\\
			\chi_1(x)&=\frac{3\sqrt3\cosh(\zeta/2)\coth\zeta}{C+2}, \qquad	&H\chi_1&=\frac94\chi_1.
		\end{aligned}
\end{equation}
The zero mode is associated with translations of the soliton, whereas the shape mode describes a localised oscillation about the static background. In particular, we have $	\lambda_1=9/4$ so that $\omega_1=\sqrt{\lambda_1}=3/2$. Choosing the cosine solution, we excite the shape mode through
\begin{equation}\label{firstordermode}
		\delta\phi_1(x,t)=\chi_1(x)\cos(\omega_1t).
\end{equation}

For later use, we recall the general solvability condition for inhomogeneous Sturm-Liouville equations, which follows from Lagrange's identity, see e.g. \cite{boyce2026elementary}. We consider
\begin{equation}\label{generalsolvability}
	(H-E)f=S, \qquad (H-E)q=0.
\end{equation}
Since $H=-\partial_x^2+U(x)$ is symmetric under formal transposition, Lagrange's identity gives
\begin{equation}\label{solvabilityidentity}
	\int_{-\infty}^{\infty}q(x)S(x)\,\dd x = \left[ f\,\partial_x q-(\partial_x f)q \right]_{-\infty}^{\infty}.
\end{equation}
If $f$ and its derivative decay sufficiently rapidly and $q$ and its derivative remain bounded, the boundary terms vanish. Consequently, a necessary condition for the existence of such a localised particular solution is
\begin{equation}\label{generalorthogonality}
	\int_{-\infty}^{\infty}q(x)S(x)\,\dd x=0.
\end{equation}
For the complex fluctuation operator, the integral is understood as a bilinear pairing without complex conjugation. This condition applies both when $q$ is a discrete eigenmode and when it is a continuum scattering solution.
	
\subsection{Second-order perturbation and continuum resonance}
	
The second-order source is quadratic in the first-order mode and consequently contains a static component and a component oscillating at twice the shape-mode frequency. We set
\begin{equation}\label{expansion}
			\phi(x,t)=\phi_s(x)+\varepsilon\chi_1(x)\cos(\omega_1t) +\varepsilon^2\left[\psi_0(x) +\psi_2(x)\cos(2\omega_1t)\right] +O(\varepsilon^3).
\end{equation}
Using $\cos^2(\omega_1t)=[1+\cos(2\omega_1t)]/2$, the second-order equation separates into
\begin{equation}\label{secondorder}
		\begin{aligned}
			H\psi_0&=S,\\
			(H-4\omega_1^2)\psi_2&=S,			
		\end{aligned}
		\qquad \qquad S(x)=-\frac14W(x)\chi_1(x)^2 =\frac{81C^2(C^3-6C^2-4)} {8(C-1)^2(C+2)^4}.
\end{equation}
Both inhomogeneous equations can be solved analytically, yielding the particular solutions
\begin{eqnarray}
	\psi_0(x)&=&-\frac9{16}\zeta\chi_0(x) -\frac9{16}\frac{11C^2+2C-16}{(C-1)(C+2)^2},  \label{psi0} \\
	\psi_2(x)&=&-\frac9{16} \frac{2C^2-7C+2}{(C-1)(C+2)^2}.  \label{g}
\end{eqnarray}
These expressions can be verified by direct substitution into equation \eqref{secondorder}. The static solution is defined up to addition of a multiple of the translational zero mode $\chi_0$. For the regularised soliton, the second-harmonic particular solution is exponentially localised
\begin{equation}\label{decay}
		\psi_2(x)=O(e^{-\sqrt3|x|}),\qquad \partial_x\psi_2(x)=O(e^{-\sqrt3|x|}) \qquad  \text{for} \,\, |x|\longrightarrow\infty.
\end{equation}
	
The continuous spectrum of $H$ was found in \cite{CorreaFringTaira} to be parametrised by
\begin{equation}\label{continuum}
		E(\kappa)=3+\kappa^2,\qquad \kappa\in\mathbb R,
\end{equation}
where $E=\lambda=\omega^2$ denotes a spectral value of the linearised operator. The second harmonic therefore satisfies
\begin{equation}\label{kappastar}
		(2\omega_1)^2=9=3+\kappa_*^2, \qquad \kappa_*=\sqrt6.
\end{equation}
Thus the second-harmonic frequency lies strictly inside the continuous spectrum, despite the existence of the exact localised particular solution \eqref{g}. Unlike a bound state in the continuum, as originally considered by von Neumann and Wigner \cite{vNW}, this solution satisfies an inhomogeneous rather than a homogeneous spectral equation. Its localisation results from the cancellation of the on-shell coupling of the nonlinear source to the scattering states, as demonstrated below.

\subsection{Third-order perturbation and frequency correction}

Next we investigate whether this unusual localisation persists at third order. At this order, the nonlinear terms generate contributions oscillating at both the fundamental frequency $\omega_1$ and its third harmonic $3\omega_1$. The former requires particular attention, since it coincides with the eigenfrequency of the shape mode. A non-vanishing projection of the forcing onto this mode produces a resonant response containing a secular term proportional to $t\sin(\omega_1 t)$. Such a term grows linearly in time and eventually invalidates the perturbative expansion, even if the underlying oscillation remains bounded.

To eliminate secular terms, we employ the Poincar\'e-Lindstedt method, see e.g.~\cite{bender2013advanced}. Introducing the rescaled time $\tau=\Omega t$, we initially allow for a frequency expansion
\begin{equation}
	\Omega=\omega_1+\varepsilon\nu_1 +\varepsilon^2\nu+O(\varepsilon^3).
\end{equation}
At second order, the perturbation equation becomes
\begin{equation}
	(\omega_1^2\partial_\tau^2+H)\delta\phi_2 =-\frac12W\chi_1^2\cos^2\tau +2\omega_1\nu_1\chi_1\cos\tau.
\end{equation}
Since $\chi_1(x)\cos\tau$ solves the corresponding homogeneous equation, we apply the general solvability condition \eqref{generalorthogonality} after projecting the forcing onto the fundamental temporal harmonic. Multiplying by $\cos\tau$ and integrating over one oscillation period eliminates the quadratic forcing, since $\int_0^{2\pi}\cos^3\tau\,\dd\tau=0$, whereas $\int_0^{2\pi}\cos^2\tau\,\dd\tau=\pi$. The resulting spatial solvability condition becomes
\begin{equation}
	2\omega_1\nu_1  I_1 =0, \qquad I_1= 	\int_{-\infty}^{\infty}\chi_1^2\,\dd x= 3 \sqrt{3},
\end{equation}
and therefore we conclude that $\nu_1=0$. Thus the frequency expansion reduces to $\Omega=\omega_1+\varepsilon^2\nu +O(\varepsilon^3)$. The frequency correction $\nu$ is determined by applying the same solvability argument at third order. A resonance at the fundamental frequency occurs unless the nonlinear forcing has vanishing projection onto the shape mode. Since $H$ is symmetric under formal transposition, the corresponding projection involves the bilinear pairing without complex conjugation. We therefore consider the expansion
\begin{equation}\label{thirdorderexpansion}
		\begin{aligned}
			\phi(x,t)=\phi_s(x)+\varepsilon\chi_1(x)\cos\tau &+\varepsilon^2\left[\psi_0(x) +\psi_2(x)\cos(2\tau)\right]\\
			&+\varepsilon^3\left[ \psi_1^{(3)}(x)\cos\tau +\psi_3(x)\cos(3\tau)\right] +O(\varepsilon^4).
		\end{aligned}
\end{equation}
Using $\cos\tau\cos(2\tau) =[\cos(3\tau)+\cos\tau]/2$ and $\cos^3\tau=[3\cos\tau+\cos(3\tau)]/4$, the third-order perturbation equation becomes
\begin{equation}\label{thirdorderequations}
		\begin{aligned}
			(H-\omega_1^2)\psi_1^{(3)} & =2\omega_1\nu\chi_1 -W\chi_1\left(\psi_0+\frac12\psi_2\right) -\frac18 Z\chi_1^3,\\
			(H-9\omega_1^2)\psi_3&=S_3,  \qquad  \qquad  \text{with} \,\, S_3(x)=-\frac12W\chi_1\psi_2 -\frac1{24}Z\chi_1^3.
		\end{aligned}
\end{equation}
Projecting the fundamental-frequency equation in \eqref{thirdorderequations} onto $\chi_1$ yields the solvability condition
\begin{equation}\label{frequencycondition}
		2\omega_1\nu I_1=I_2, \qquad I_2=\int_{-\infty}^{\infty} \left[ W\chi_1^2\left(\psi_0+\frac12\psi_2\right) +\frac18Z\chi_1^4 \right]\dd x =\frac{81\sqrt3}{32}.
\end{equation}
Therefore we obtain
\begin{equation}\label{frequencyshift}
	\nu=\frac9{32},\qquad \Omega=\frac32+\frac9{32}\varepsilon^2 +O(\varepsilon^3).
\end{equation}
Without the frequency correction, the third-order solution would contain the secular contribution $-9\varepsilon^3t\chi_1(x)\sin(\omega_1t)/32$. This term is absorbed into the oscillation with the corrected frequency $\Omega$.
	
With this frequency correction, both third-order inhomogeneous equations admit explicit localised particular solutions. One convenient choice for the fundamental-frequency correction is
\begin{equation}\label{psi13}
			\psi_1^{(3)}(x)={} -\frac9{16\sqrt3}\,\zeta\,\partial_x\chi_1(x) +\frac{27\sqrt3}{64} \frac{15C^3-10C^2-22C+20} {\sinh(\zeta/2)(C-1)(C+2)^3},
\end{equation}
while the third-harmonic correction is
\begin{equation}\label{psi3}
		\psi_3(x)=\frac{9\sqrt3}{64} \frac{3C^3-10C^2+14C-4} {\sinh(\zeta/2)(C-1)(C+2)^3}.
\end{equation}
Both expressions satisfy their respective equations in~\eqref{thirdorderequations}, as can be verified by direct substitution. The solution for $\psi_1^{(3)}$ is defined up to a multiple of the homogeneous shape mode $\chi_1$. Here $\partial_x=\sqrt3\,\partial_\zeta$ is understood when differentiating the explicit profile $\chi_1(x)$.
	
In particular, the third-harmonic correction and its spatial derivative decay exponentially,
\begin{equation}\label{thirddecay}
		\psi_3(x)=O\left(e^{-3\sqrt3|x|/2}\right),\qquad \partial_x\psi_3(x) =O\left(e^{-3\sqrt3|x|/2}\right), \qquad  \text{for} \,\, |x|\longrightarrow\infty.
\end{equation}
Its squared frequency also belongs to the continuum,
\begin{equation}\label{kappa3}
		(3\omega_1)^2=\frac{81}{4} =3+\kappa_3^2,\qquad \kappa_3=\frac{\sqrt{69}}2.
\end{equation}
We have therefore found localised particular responses at both the second and third harmonics, even though both frequencies fall inside the continuous spectrum. 

\begin{figure}[h]
	\begin{minipage}[b]{\textwidth}      
		\centering
		\includegraphics[width=0.44\textwidth]{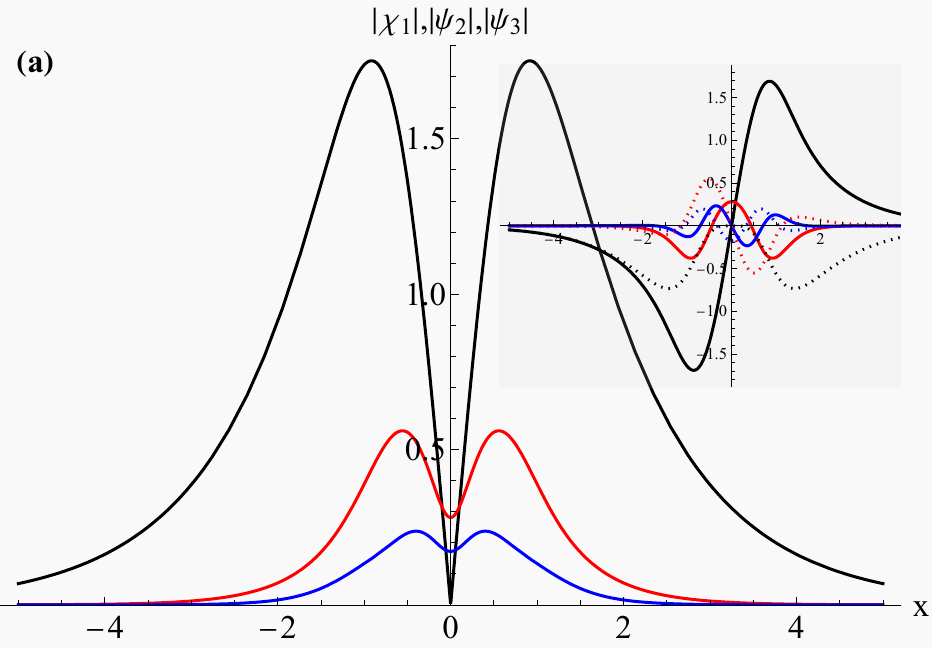}
		\includegraphics[width=0.55\textwidth]{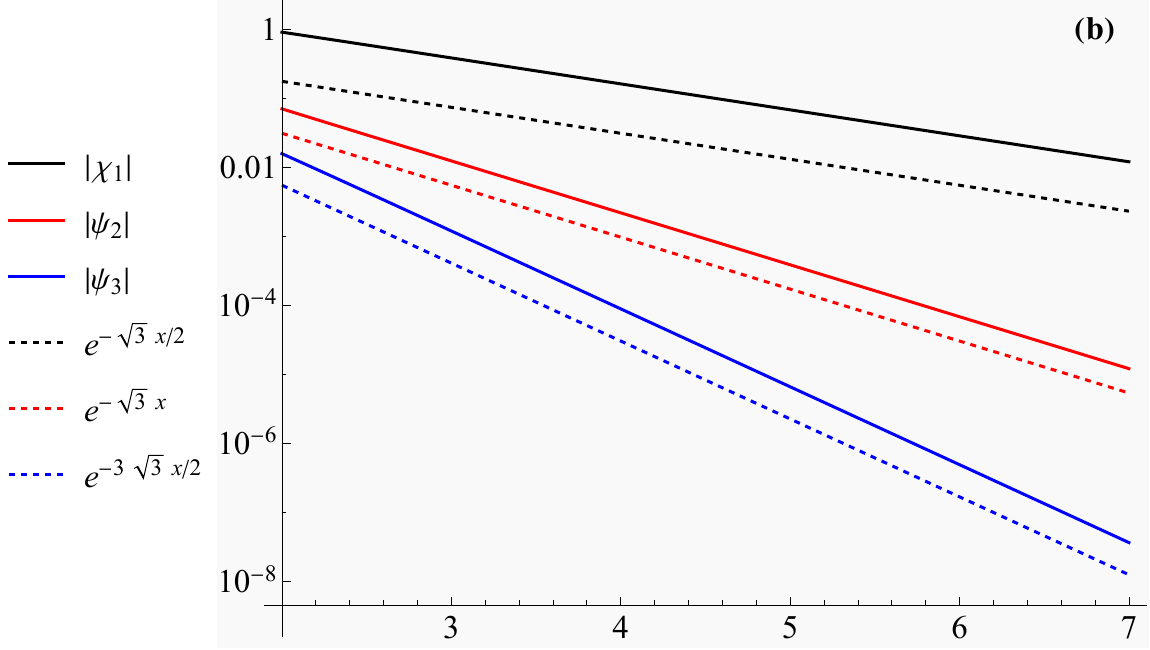}
	\end{minipage}   
	\caption{Spatial localisation of the shape mode and its nonlinear harmonic corrections for $\beta=i\pi/2$. (a) Absolute values of the shape mode $\chi_1(x)$ and the second- and third-harmonic particular solutions $\psi_2(x)$ and $\psi_3(x)$, respectively. The inset displays their real (solid) and imaginary (dotted) parts. (b) Semilogarithmic representation of the corresponding profiles for positive $x$, illustrating their exponential decay. The second- and third-harmonic corrections decay asymptotically as $e^{-\sqrt3 x}$ and $e^{-3\sqrt3 x/2}$, respectively, as indicated by the dashed reference lines.}
	\label{spprofiles}
\end{figure}

The spatial profiles of the shape mode and its second- and third-harmonic corrections are shown in figure \ref{spprofiles} for $\beta=i\pi/2$. The absolute values and the real and imaginary parts illustrate the spatial localisation of these complex-valued solutions. The semilogarithmic plot further confirms the exponential decay of the profiles, in agreement with the asymptotic behaviour derived above. In particular, both higher-harmonic corrections remain localised despite their frequencies lying inside the continuous spectrum.

In the next section we examine the corresponding on-shell couplings to the scattering states, beginning with the second harmonic.

\section{Exact cancellation of the on-shell continuum amplitudes}

We now apply the general solvability condition \eqref{generalorthogonality} to the second- and third-harmonic equations. In contrast to the fundamental-frequency resonance considered above, the relevant homogeneous solutions are continuum scattering states. The corresponding overlap integrals determine the on-shell coupling of the nonlinear sources to the radiation channels.

\subsection{Second-harmonic cancellation}

The quadratic source $S(x)$ in \eqref{secondorder} drives the second harmonic at frequency $2\omega_1=3$, corresponding to $E=9$ and $\kappa_*=\sqrt6$. For the subsequent analysis, it is useful to introduce the continuum scattering solutions at a general positive wavenumber $\kappa$. Defining
\begin{equation}\label{scatteringcoefficients}
		a_\kappa(\zeta)
		=\frac32-\kappa^2-\frac9{2(C+2)}, \qquad
		b_\kappa(\zeta)
		=\frac{3\sqrt3\,\kappa C\coth(\zeta/2)}
		{2(C+2)},
		\qquad
		\theta_\kappa=\frac{\kappa\zeta}{\sqrt3},
\end{equation}
with $C$ and $\zeta$ as defined above, two independent solutions of $(H-3-\kappa^2)q=0$ are
\begin{eqnarray}
	q_e(x;\kappa) &=&\frac{b_\kappa\cos\theta_\kappa -a_\kappa\sin\theta_\kappa}{\kappa}, \label{qeven}\\
	q_o(x;\kappa) &=&\frac{2\sqrt3}{3+4\kappa^2} \left(a_\kappa\cos\theta_\kappa +b_\kappa\sin\theta_\kappa\right). \label{qodd}
\end{eqnarray}
These expressions are normalised consistently with the scattering solutions used in \cite{CorreaFringTaira}. Their Wronskian is
\begin{equation}
	\mathcal W_x(q_e,q_o) =q_e\partial_xq_o-(\partial_xq_e)q_o =\frac{\sqrt3}{2}(\kappa^2+3),
\end{equation}
which confirms their linear independence for every real $\kappa>0$.

At the second-harmonic wavenumber $\kappa_*=\sqrt6$, they simplify considerably. Introducing
\begin{equation}\label{AB}
	A(\zeta)=\frac{C\coth(\zeta/2)}{C+2}, \qquad B(\zeta)=\frac{C+3}{C+2}, \qquad \theta=\sqrt2\,\zeta,
\end{equation}
we have
\begin{equation}
	a_{\kappa_*}=-\frac92B, \qquad b_{\kappa_*}=\frac{9\sqrt2}{2}A, \qquad \theta_{\kappa_*}=\theta.
\end{equation}
The corresponding scattering functions $q_j(x)=q_j(x;\kappa_*)$ therefore satisfy
\begin{equation}\label{homogeneous}
	(H-9)q_e=0,\qquad (H-9)q_o=0.
\end{equation}

The regularised potential $U(x)$ is generally complex, so $H$ is not Hermitian in the standard $L^2$ inner product. The adjoint scattering solutions enter the corresponding sesquilinear projections. Since the differential expression for $H$ is symmetric under formal transposition, these projections can equivalently be written as bilinear integrals with $q_e$ and $q_o$, without complex conjugation. We define
\begin{equation}\label{Mj}
	\mathcal M_j^{(2)}=\int_{-\infty}^{\infty}
	q_j(x)S(x)\,\dd x,\qquad j\in\{e,o\}.
\end{equation}
The vanishing of both amplitudes means that the quadratic source has no on-shell coupling to either independent scattering channel. Indeed we find:

\emph{For a nonsingular shifted type-I soliton, both on-shell overlaps vanish exactly:}
\begin{equation}\label{zero}
	\mathcal M_e^{(2)}=\mathcal M_o^{(2)}=0.
\end{equation}

To see this, recall from equation~\eqref{g} that $(H-9)\psi_2=S$, while $(H-9)q_j=0$. Subtracting the latter equation, multiplied by $\psi_2$, from the former equation, multiplied by $q_j$, and using $H=-\partial_x^2+U(x)$, we obtain Lagrange's identity for this Sturm-Liouville operator
\begin{equation}\label{green}
	\begin{aligned}
		\int_{-\infty}^{\infty}q_jS\,\dd x
		&=\int_{-\infty}^{\infty}
		[q_j(H-9)\psi_2-\psi_2(H-9)q_j] \,\dd x\\
		&=\int_{-\infty}^{\infty}
		[\psi_2\,\partial_x^2q_j-q_j\,\partial_x^2\psi_2]\,\dd x\\
		&=\int_{-\infty}^{\infty}\partial_x
		[\psi_2\,\partial_xq_j-(\partial_x\psi_2)q_j]\,\dd x\\
		&=\left[\psi_2\,\partial_xq_j
		-(\partial_x\psi_2)q_j\right]_{-\infty}^{\infty}=0.
	\end{aligned}
\end{equation}
The potential and spectral terms have cancelled identically. The remaining integrand is a total derivative of the Wronskian $	\mathcal W_x(\psi_2,q_j)$. The scattering functions and their first spatial derivatives remain bounded along the admissible shifted contour, whereas $\psi_2$ and $\partial_x\psi_2$ decay exponentially, by equation~\eqref{decay}. Both boundary terms therefore vanish.

For completeness, the Wronskians can be evaluated explicitly. Keeping in mind that $\partial_x=\sqrt3\,\partial_\zeta$, we use the expressions from equations~\eqref{qeven}, \eqref{qodd} and \eqref{g} to compute
\begin{eqnarray}
		\mathcal W_x(\psi_2,q_e) &=&\frac92\left[\psi_2(A_\zeta+B) -(\partial_\zeta\psi_2)A\right]\cos\theta +\frac{9\sqrt2}{4}\left[\psi_2(B_\zeta-2A) -(\partial_\zeta\psi_2)B\right]\sin\theta,  \qquad \\
		\mathcal W_x(\psi_2,q_o) &=&\left[\psi_2(2A-B_\zeta) +(\partial_\zeta\psi_2)B\right]\cos\theta +\sqrt2\left[\psi_2(A_\zeta+B) -(\partial_\zeta\psi_2)A\right]\sin\theta,
\end{eqnarray}
where
\begin{equation}\label{ABderivatives}
		A_\zeta=\frac{C^2-2C-2}{(C-1)(C+2)^2},\quad
		B_\zeta=-\frac{\sinh\zeta}{(C+2)^2}, \quad
		\partial_\zeta\psi_2
		=\frac{9\sinh\zeta(2C^3-18C^2+21C-14)}
		{16(C-1)^2(C+2)^3}.
\end{equation}
All these quantities are understood as functions of $x$ through $\zeta=\beta+\sqrt3x$. Direct differentiation verifies
\begin{equation}\label{totalderivatives}
		\partial_x\mathcal W_x(\psi_2,q_j)=q_jS,\qquad 
		\lim_{x\to\pm\infty}\mathcal W_x(\psi_2,q_j)=0,
		\qquad j\in\{e,o\}.
\end{equation}
Thus both on-shell overlaps vanish separately.

For a nearby real continuum wavenumbers, Lagrange identity gives
\begin{equation}\label{offresonant}
	\mathcal M_j^{(2)}(\kappa) =(\kappa^2-6)\int_{-\infty}^{\infty} q_j(x;\kappa)\psi_2(x)\,\dd x.
\end{equation}
The integral on the right is finite near $\kappa=\sqrt6$, so the overlap has an exact zero at the second-harmonic wavenumber.

\subsection{Third-harmonic cancellation}

The same argument applies at third order. As shown above, the third-harmonic equation \eqref{thirdorderequations} admits the exponentially localised particular solution~\eqref{psi3}, even though its frequency lies inside the continuous spectrum, as established in~\eqref{kappa3}.

At $E_3=9\omega_1^2=81/4$, the corresponding wavenumber is $\kappa_3=\sqrt{69}/2$. We denote by $q_j^{(3)}(x)=q_j(x;\kappa_3)$, $j\in\{e,o\}$, the scattering solutions defined in \eqref{qeven} and \eqref{qodd}, which satisfy $(H-E_3)q_j^{(3)}=0$. The corresponding on-shell overlaps are
\begin{equation}\label{Mjthird}
	\mathcal M_j^{(3)} =\int_{-\infty}^{\infty} q_j^{(3)}(x)S_3(x)\,\dd x, \qquad j\in\{e,o\}.
\end{equation}
Applying the general solvability identity \eqref{solvabilityidentity} with $f=\psi_3$ and $q=q_j^{(3)}$, we obtain
\begin{equation}\label{greenthird}
		\mathcal M_j^{(3)} =\left[ \psi_3\,\partial_xq_j^{(3)} -(\partial_x\psi_3)q_j^{(3)} \right]_{-\infty}^{\infty}=0.
\end{equation}
The boundary terms vanish because $\psi_3$ and its derivative decay exponentially, whereas the scattering solutions and their derivatives remain bounded. Consequently, we have established the third-harmonic counterpart of the second-order cancellation~\eqref{zero}.

The dependence on the continuum wavenumber can again be made explicit. For scattering solutions satisfying $(H-3-\kappa^2)q_j(x;\kappa)=0$, Lagrange's identity gives
\begin{equation}\label{offresonantthird}
	\mathcal M_j^{(3)}(\kappa) =\left(\kappa^2-\frac{69}{4}\right) \int_{-\infty}^{\infty} q_j(x;\kappa)\psi_3(x)\,\dd x.
\end{equation}
The integral is finite near $\kappa_3=\sqrt{69}/2$, so the overlap vanishes exactly at the third-harmonic wavenumber, just as its second-harmonic counterpart does at $\kappa_*=\sqrt6$.

These cancellations concern the specific nonlinear sources generated by the perturbative shape-mode excitation. They do not exclude homogeneous scattering contributions, transient radiation from other initial conditions, or radiation at higher perturbative orders.

The cancellations also extend to the negative type-I branch, obtained by replacing $\beta$ with $\beta+i\pi$. For $\beta=3i\pi/2$, we have $C=-i\sinh(\sqrt3\,x)$, and the corresponding background and perturbative profiles remain nonsingular along the real spatial axis. Choosing a continuous logarithm, the soliton now approaches the equivalent vacuum values $0$ and $-2\pi i$ at the two spatial infinities. Since the second- and third-harmonic particular solutions remain exponentially localised, Lagrange's identity again implies the exact vanishing of their on-shell continuum overlaps.

\begin{figure}[h]
	\begin{minipage}[b]{\textwidth}      
		\centering
		\includegraphics[width=0.52\textwidth]{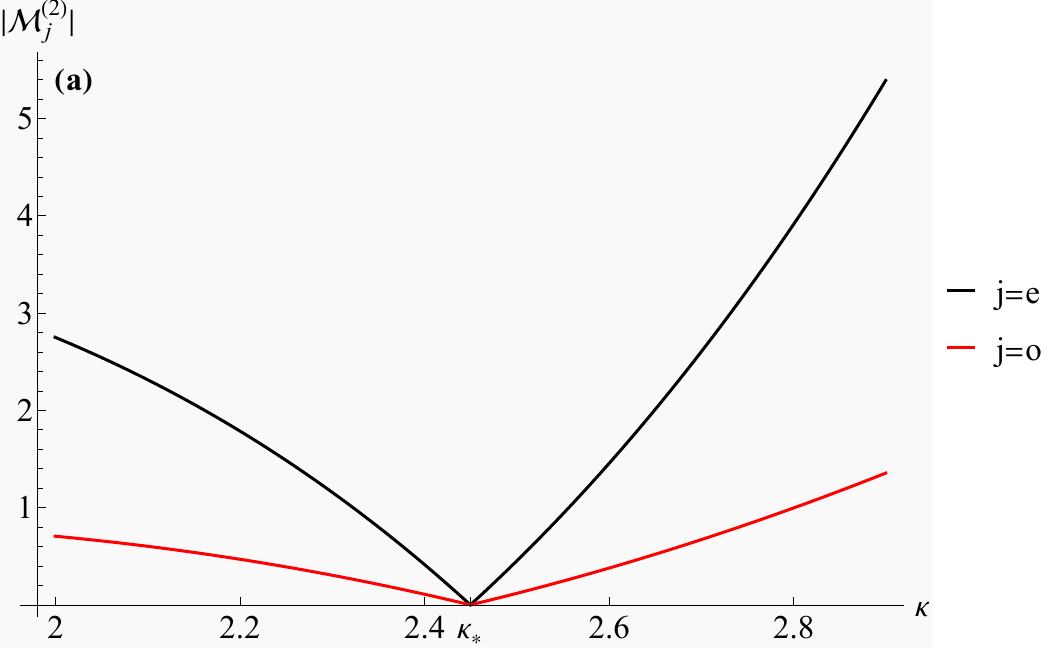}
		\includegraphics[width=0.46\textwidth]{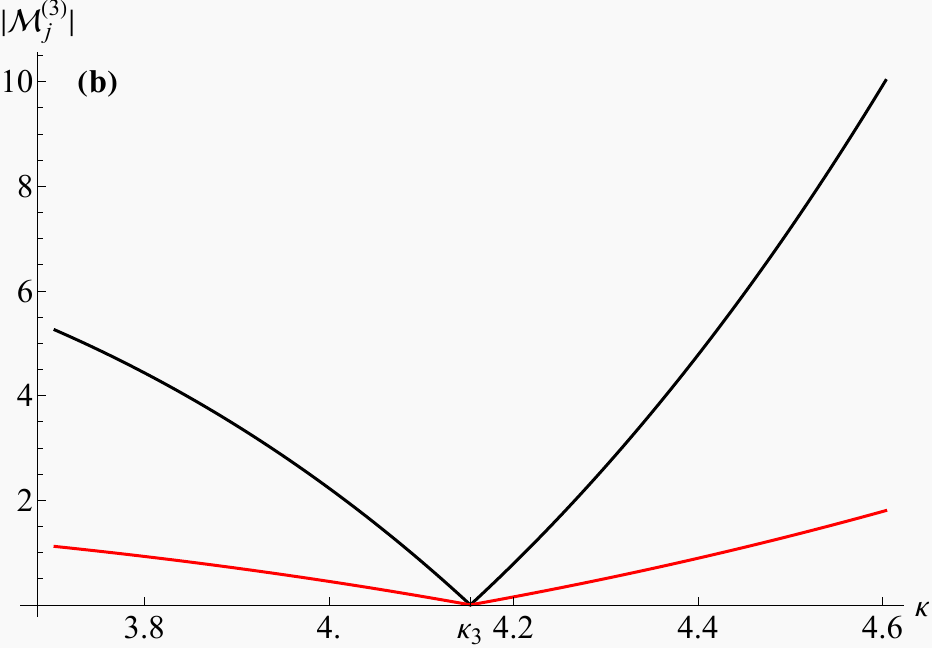}
	\end{minipage}   
	\caption{Magnitudes of the continuum overlap integrals $\mathcal M_j^{(n)}(\kappa)$ for the second (a) and third (b) harmonics, evaluated for the complex shift $\beta=i\pi/2$. The black and red curves correspond to the even- and odd-labelled scattering channels, respectively. Both overlaps vanish at the resonant wavenumbers $\kappa_*=\sqrt6$ in (a) and $\kappa_3=\sqrt{69}/2$ in (b), in agreement with the analytical results.}
	\label{overlaps}
\end{figure}

The dependence of the overlap integrals on the continuum wavenumber is illustrated in figure \ref{overlaps}. Both scattering channels exhibit zeros precisely at the second- and third-harmonic resonances, $\kappa_*=\sqrt6$ and $\kappa_3=\sqrt{69}/2$, respectively. Away from these values the overlaps are generally nonzero, demonstrating that the cancellation is specific to the resonant wavenumbers rather than a general decoupling of the nonlinear sources from the continuum.

While these identities establish the exact cancellation of the on-shell radiation amplitudes, they do not explain why the nonlinear sources admit such localised particular solutions. In the following section, we investigate whether this property can be understood in terms of the higher commuting symmetries of the integrable Bullough-Dodd equation.

\section{Higher-symmetry selection rules for nonlinear radiation}
\label{subsec:higherselection}

The preceding cancellations follow from the existence of exponentially localised particular solutions. We now investigate why the nonlinear sources possess the required vanishing continuum projections. Being an integrable system, the Bullough-Dodd equation has higher generalised symmetries whose flows commute with the physical evolution. Their spectral eigenvalues provide additional labels for the shape mode and continuum states, and hence impose restrictions on nonlinear mode coupling. We establish these restrictions first at the second harmonic and then at the third.

\subsection{The fifth-order symmetry and the static soliton}

It is convenient to introduce light-cone coordinates and their derivatives,
\begin{equation}\label{lightcone}
	x_+=\frac{t+x}{2},\qquad x_-=\frac{t-x}{2},\qquad D_\pm=\partial_t\pm\partial_x=\frac{\partial}{\partial x_\pm},
\end{equation}
in which the field equation becomes
\begin{equation}\label{BDlightcone}
	D_+D_-\phi=f(\phi),\qquad f(\phi)=-e^\phi+e^{-2\phi}.
\end{equation}
The exponential nonlinearity satisfies the identity
\begin{equation}\label{fidentity}
	f''(\phi)+f'(\phi)-2f(\phi)=0,
\end{equation}
where primes denote differentiation with respect to $\phi$. This relation will be useful in verifying the higher conservation law. The connection between the Bullough-Dodd nonlinearity and the existence of higher symmetries is well established \cite{zhiber1979klein,kaptsov2018int}. An explicit higher conservation law and its associated conserved charge are presented in Appendix~\ref{app:BDcharge}.

Following \cite{kaptsov2018int}, we abbreviate $u_n=D_+^n\phi$ and recall that the Bullough-Dodd equation admits the fifth-order generalised symmetry
\begin{equation}\label{G5higher}
	\mathcal G_5[\phi]
	=u_5+5\big(u_2u_3-u_1^2u_3-u_1u_2^2\big)+u_1^5.
\end{equation}
Here, a generalised symmetry is understood as an infinitesimal transformation that preserves the equation of motion and whose characteristic may depend on higher derivatives of the field. More precisely, under the transformation $\phi\mapsto\phi+\alpha\mathcal G_5[\phi]$, the Bullough-Dodd equation remains satisfied to first order in $\alpha$, provided
\begin{equation}\label{G5symmetry}
	D_+D_-\mathcal G_5=f'(\phi)\mathcal G_5
\end{equation}
holds on solutions of \eqref{BDlightcone}. This identity, which is the linearised Bullough-Dodd equation evaluated on $\mathcal G_5$, can be verified by direct substitution.

The resulting infinitesimal symmetry is compatible with physical time evolution. In particular, when expressed as evolutionary vector fields on the space of Cauchy data, the higher symmetry and the physical evolution commute. This commutativity will be used below to derive constraints on nonlinear mode coupling. No particular Poisson structure is assumed, and we do not identify the symmetry with the Hamiltonian flow of the conserved charge discussed in Appendix~\ref{app:BDcharge}.

Next we examine the action of this higher symmetry on the static type-I soliton $\phi_s$. Since $D_+=\partial_x$ when acting on a static field, we may use the static equation and its first integral. The latter follows by multiplying the static equation by $\phi_s'$ and using $\phi_s'\to0$ and $e^{\phi_s},e^{-2\phi_s}\to1$ at spatial infinity
\begin{equation}\label{staticfirstintegral}
	\phi_s''=e^{\phi_s}-e^{-2\phi_s}, \qquad \frac12(\phi_s')^2 =e^{\phi_s}+\frac12e^{-2\phi_s}-\frac32.
\end{equation}
Substituting these relations into \eqref{G5higher}, we obtain
\begin{equation}\label{G5static}
	\mathcal G_5[\phi_s]=9\,\partial_x\phi_s.
\end{equation}
Thus, on the static soliton, the higher symmetry reduces to an infinitesimal translation rather than leaving the background invariant. Since $D_+\phi$ is itself the characteristic of a commuting translation symmetry, we may introduce the modified characteristic
\begin{equation}\label{Kflow}
	\mathcal K[\phi]=\mathcal G_5[\phi]-9D_+\phi,	\qquad \mathcal K[\phi_s]=0.
\end{equation}
The modified symmetry therefore leaves the static soliton invariant while retaining its commutativity with physical time evolution. This allows us to linearise both transformations about the same background and compare their action on the discrete shape mode and the continuum scattering states.

\subsection{Higher-flow eigenvalues and continuum channels}

Having constructed a higher symmetry that leaves the static soliton invariant, we now investigate its action on the linearised fluctuations. Since the modified symmetry $\mathcal K$ commutes with physical time evolution and satisfies $\mathcal K[\phi_s]=0$, its linearisation about $\phi_s$ preserves the solutions of the linearised Bullough-Dodd equation. The discrete shape mode and the continuum scattering states can therefore be characterised not only by their physical frequencies, but also by the action of the higher symmetry. As we shall see, comparing these additional eigenvalues provides the basis for the nonlinear selection rules.

To this end, we expand the symmetry characteristic about the static background
\begin{equation}
	\mathcal K[\phi_s+\varepsilon v] =\varepsilon\mathcal K_1v+O(\varepsilon^2),
\end{equation}
where $\mathcal K_1$ denotes the linearised higher symmetry. Using $u_n^{s}=D_+^n\phi_s$, direct linearisation of \eqref{Kflow} gives
\begin{eqnarray}
		\mathcal K_1v&=&D_+^5v +5\big[u_2^s-(u_1^s)^2\big]D_+^3v +5\big[u_3^s-2u_1^su_2^s\big]D_+^2v   \label{Klinearised} \\&& \qquad \qquad \qquad \qquad \qquad \qquad \qquad+\big[5\big((u_1^s)^4 -2u_1^su_3^s-(u_2^s)^2\big)-9\big]D_+v. \notag
\end{eqnarray}

We first consider the discrete shape mode. In section 2, its time dependence was chosen as $\cos(\omega_1t)$. For the higher-symmetry analysis, it is convenient to separate the positive- and negative-frequency components of this oscillation and consider
\begin{equation}\label{shapehighermode}
	v_1(x,t)=\chi_1(x)e^{-i\omega_1t}, \qquad \omega_1=\frac32.
\end{equation}
Substituting this expression into \eqref{Klinearised}, using the explicit background and shape-mode profiles, yields
\begin{equation}\label{shapehighermu}
	\mathcal K_1v_1=\mu_1v_1, \qquad \mu_1=27i.
\end{equation}
Thus the positive-frequency shape-mode component is an eigenfunction of both the linearised physical time evolution and the higher symmetry, with the latter characterised by the additional eigenvalue $\mu_1$.

Next we consider the continuum scattering states. Unlike the spatially localised shape mode, these states approach oscillatory plane waves at spatial infinity. Although the static soliton approaches different equivalent vacuum values at the two spatial infinities, its derivatives vanish and $e^{\phi_s},e^{-2\phi_s}\to1$. Consequently, the linearised higher symmetry approaches $\mathcal K_1\to D_+^5-9D_+$. For an asymptotic plane wave $v_\kappa=e^{i(\kappa x-\omega t)}$, with $\omega^2=3+\kappa^2$, we have $D_+v_\kappa=i(\kappa-\omega)v_\kappa$. The corresponding higher-symmetry eigenvalue is therefore
\begin{equation}\label{mucontinuum}
	\mu(\kappa,\omega)
	=i\big[(\kappa-\omega)^5
	-9(\kappa-\omega)\big].
\end{equation}
Thus, in addition to their physical frequency $\omega$, the asymptotic scattering states are characterised by an eigenvalue of the fifth-order symmetry.

Of particular interest are the continuum states whose physical frequencies coincide with the second and third harmonics of the shape mode. At the second harmonic, $\omega=2\omega_1=3$ and $\kappa=\pm\sqrt6$, giving
\begin{equation}\label{musecond}
	\mu_\pm^{(2)} =i\big(-2376\pm972\sqrt6\big), \qquad \mu_\pm^{(2)}\ne2\mu_1=54i.
\end{equation}
Similarly, at the third harmonic, $\omega=3\omega_1=9/2$ and $\kappa=\pm\sqrt{69}/2$, we obtain
\begin{equation}\label{muthird}
	\mu_\pm^{(3)} =i\big(-24219\pm2916\sqrt{69}\big), \qquad \mu_\pm^{(3)}\ne3\mu_1=81i.
\end{equation}
Hence, although the second and third harmonics are resonant with continuum states under physical time evolution, their higher-symmetry eigenvalues do not match the corresponding multiples of $\mu_1$. This mismatch will provide the basis for the nonlinear selection rules derived below.

\subsection{The dual scattering states}

The asymptotic calculation above identifies the higher-symmetry eigenvalues associated with the two continuum momenta. To derive constraints on the radiation amplitudes, however, we must establish how the higher symmetry acts on the corresponding dual scattering states. Recall that the on-shell amplitudes in section~3 involve bilinear overlaps of the form $\int q(x)S(x)\,\dd x$, without complex conjugation. In the phase-space formulation, these projections involve left eigenvectors of the linearised generators, or equivalently eigenvectors of their formal transposes.

To describe fluctuations about the static soliton, we write $\phi=\phi_s+\eta$, where $\eta(x,t)$ denotes an arbitrary perturbation of the background. Since $\phi_s$ is time independent, the corresponding Cauchy data are
\begin{equation}
	\mathbf a=
	\begin{pmatrix}
		\eta\\\pi
	\end{pmatrix},
	\qquad
	\pi=\partial_t\eta=\partial_t\phi.
\end{equation}
This choice places the static soliton at $\mathbf a=0$. Moreover, since both the physical time evolution and the modified higher symmetry leave the static soliton invariant, they have a common fixed point about which their generators can be linearised.

For the physical time evolution, the linearised Bullough-Dodd equation takes the first-order form
\begin{equation}
	\partial_t\mathbf a=A\mathbf a,
	\qquad
	A=
	\begin{pmatrix}
		0&1\\
		-H&0
	\end{pmatrix}.
\end{equation}
Similarly, the linearised higher symmetry $\mathcal K_1$ defines a generator $B$ acting on the same Cauchy data. Its first component is $\mathcal K_1\eta$, while the second is obtained by differentiating this expression with respect to time and eliminating higher time derivatives using the linearised field equation. Since the underlying nonlinear symmetries commute and share the same fixed point, their linearised generators satisfy
\begin{equation}
	[A,B]=0.
\end{equation}

We next relate these generators to the scattering solutions used previously. At a continuum energy $E=3+\kappa^2$, the spatial fluctuation equation is
\begin{equation}
	(H-3-\kappa^2)q=0.
\end{equation}
The solutions $q_e$ and $q_o$ introduced in section~3 span its two-dimensional solution space. For the present purpose, it is convenient to choose instead linear combinations $q_\pm$ with definite asymptotic momenta,
\begin{equation}
	q_\pm(x;\kappa)\sim e^{\pm i\kappa x}, \qquad x\to+\infty.
\end{equation}
These are the corresponding Jost solutions. They describe the same continuum channels as $q_e$ and $q_o$, but their asymptotic normalisation is better suited to the higher-symmetry analysis. We shall need this construction only at the second- and third-harmonic wavenumbers $\kappa_2=\sqrt6$ and $\kappa_3=\sqrt{69}/2$.

To construct the dual continuum states, we use the bilinear spatial pairing
\begin{equation}\label{phasepairing}
	\langle\mathbf b,\mathbf a\rangle =\int_{-\infty}^{\infty} [b_1(x)a_1(x)+b_2(x)a_2(x)]\,\mathrm dx,
\end{equation}
without complex conjugation. The corresponding formal transpose is defined by transferring spatial derivatives between the two entries through integration by parts. In particular, $D^{\mathrm T}=-D$, where $D=\partial_x$. When one entry is a continuum scattering state, the pairing will be used with spatially localised sources, for which the integrals converge and the boundary terms vanish.

With respect to this pairing, the linearised
physical generator has the formal transpose
\begin{equation}
	A^{\mathrm T}=
	\begin{pmatrix}
		0&-H\\
		1&0
	\end{pmatrix}.
\end{equation}
 Abbreviating $\sigma=-i\omega$, with $\omega^2=3+\kappa^2$, we construct the dual phase-space vectors
\begin{equation}
	\widetilde{\mathbf v}_\pm=
	\begin{pmatrix}
		\sigma q_\pm\\q_\pm
	\end{pmatrix}.
\end{equation}
Because $Hq_\pm=\omega^2q_\pm$ and $\sigma^2=-\omega^2$, direct substitution gives
\begin{equation}
	A^{\mathrm T}\widetilde{\mathbf v}_\pm	=\sigma\widetilde{\mathbf v}_\pm.
\end{equation}
Thus the dual Jost states are eigenvectors of the transposed physical generator. We must now establish the corresponding property for the higher-symmetry generator.

To determine its eigenvalues, we examine the asymptotic behaviour of the linearised generators. Since $e^{\phi_s},e^{-2\phi_s}\to1$ at both spatial infinities, we have $H\to-D^2+3$, where $D=\partial_x$. The physical generator therefore reduces to
\begin{equation}
	A_\infty=
	\begin{pmatrix}
		0&1\\
		D^2-3&0
	\end{pmatrix}.
\end{equation}
The corresponding vacuum fluctuation equation is
\begin{equation}
	\eta_{tt}-\eta_{xx}+3\eta=0.
\end{equation}

In the same limit, the scalar higher-symmetry operator becomes $\mathcal K_1=D_+^5-9D_+$, where $D_+=\partial_t+D$. Using the vacuum equation to eliminate second and higher time derivatives, we can write
\begin{equation}
	\mathcal K_1\eta =p_5(D)\eta+r_4(D)\pi,
\end{equation}
with differential polynomials
\begin{equation}\label{Binfty}
		p_5(D)=16D^5-60D^3+36D, \qquad \quad	r_4(D)=16D^4-36D^2.
\end{equation}
Since the second phase-space component is obtained by differentiating the first with respect to time, we find
\begin{equation}
	B_\infty=p_5(D)I+r_4(D)A_\infty.
\end{equation}
This is the vacuum limit of the linearised higher-symmetry generator $B$, expressed entirely in terms of spatial derivatives and the physical time generator.

To determine the action on dual states, we take the formal transpose. Integration by parts gives $D^{\mathrm T}=-D$. Since $p_5$ is odd and $r_4$ is even in $D$, we obtain
\begin{equation}
	B_\infty^{\mathrm T}
	=-p_5(D)I+r_4(D)A_\infty^{\mathrm T}.
\end{equation}
Acting on the asymptotic forms of the dual Jost vectors therefore yields
\begin{equation}\label{dualvacuumvalues}
		\widetilde\mu_\pm(\kappa,\omega) =-p_5(\pm i\kappa) +\sigma r_4(\pm i\kappa) =\mu(\mp\kappa,\omega).
\end{equation}
The reversal of the momentum label is a consequence of transposing the odd spatial derivatives. Thus the two dual channels have the same set of higher-symmetry eigenvalues obtained previously for the ordinary continuum modes.

It remains to be shown that these asymptotic eigenvalue relations extend to the complete Jost solutions. Because $[A,B]=0$, their formal transposes also commute $[A^{\mathrm T},B^{\mathrm T}]=0$. Consequently, the action of $B^{\mathrm T}$ preserves the solution space of the physical eigenvalue equation at fixed $\sigma$.

Thus, considering the vector
\begin{equation}
	\mathbf d_\pm= (B^{\mathrm T}-\widetilde\mu_\pm) \widetilde{\mathbf v}_\pm
\end{equation}
the commutativity immediately gives
\begin{equation}
	(A^{\mathrm T}-\sigma) \mathbf d_\pm=0.
\end{equation}
Furthermore, the asymptotic relations derived above imply that $\mathbf d_\pm$ has vanishing plane-wave coefficients as $x\to+\infty$. The spatial component of this homogeneous equation is the same second-order scattering equation considered in section~3. At nonzero real $\kappa$, uniqueness of its Jost solutions therefore implies $\mathbf d_\pm=0$, provided the standard differentiable Jost asymptotics hold for the higher-order operator.

We conclude that
\begin{equation}
	B^{\mathrm T}\widetilde{\mathbf v}_\pm	=\widetilde\mu_\pm
	\widetilde{\mathbf v}_\pm.
\end{equation}
Hence the dual Jost states are simultaneous eigenvectors of the transposed physical and higher-symmetry generators. This is the property required for the nonlinear selection rules derived below. Since the Jost states and the solutions $q_e,q_o$ span the same scattering solution space, vanishing projections onto both Jost channels imply the vanishing of the corresponding on-shell overlaps in the basis used in section~3.

\subsection{Quadratic selection rule}

Having established the simultaneous eigenvalue structure of the shape mode and the dual continuum states, we now examine its consequences for nonlinear mode coupling. The essential observation is that commutativity of the physical and higher-symmetry evolutions imposes constraints not only on their linearised generators, but also on their nonlinear interactions.

To make these constraints explicit, we extend the linearised generators $A$ and $B$ introduced above to the corresponding nonlinear vector fields $X$ and $Y$ on the space of Cauchy data. The physical evolution is governed by
\begin{equation}
	\partial_t\mathbf a=X(\mathbf a),
\end{equation}
where
\begin{equation}\label{Xdefinition}
	X(\mathbf a)=
	\begin{pmatrix}
		\pi\\
		\eta_{xx}+f(\phi_s+\eta)-f(\phi_s)
	\end{pmatrix},
\end{equation}
with $f(\phi)$ defined in \eqref{BDlightcone}. We have used the static field equation for $\phi_s$.

The modified fifth-order symmetry acts on the same Cauchy data through the vector field
\begin{equation}\label{Ydefinition}
	Y(\mathbf a)=
	\left.
	\begin{pmatrix}
		\mathcal K[\phi]\\
		\partial_t\mathcal K[\phi]
	\end{pmatrix}
	\right|_{\phi=\phi_s+\eta,\;\partial_t\phi=\pi}.
\end{equation}
The second component describes the induced transformation of $\pi=\partial_t\phi$. Higher time derivatives are eliminated using the field equation and its differential consequences, so that $Y$ depends only on the Cauchy data and their spatial derivatives.

As the modified higher symmetry commutes with physical time evolution, the vector fields $X$ and $Y$ commute. We shall exploit this property by comparing their expansions at successive orders.

Since both vector fields vanish at $\mathbf a=0$, their expansions about this common fixed point take the form
\begin{equation}\label{flowexpansions}
	\begin{aligned}
		X(\mathbf a)
		&=A\mathbf a+\frac12N(\mathbf a,\mathbf a)
		+O(\mathbf a^3),\\
		Y(\mathbf a)
		&=B\mathbf a+\frac12P(\mathbf a,\mathbf a)
		+O(\mathbf a^3).
	\end{aligned}
\end{equation}
Here $A$ and $B$ are the linearised generators, while $N$ and $P$ are symmetric bilinear maps acting on pairs of perturbations. In the expansions above, both arguments are taken to be the same perturbation $\mathbf a$. The factors $1/2$ follow from the usual Taylor-expansion convention.

For the physical evolution, expanding the Bullough-Dodd nonlinearity to second order gives
\begin{equation}
	X(\mathbf a)=A\mathbf a+
	\frac12
	\begin{pmatrix}
		0\\-W\eta^2
	\end{pmatrix}
	+O(\mathbf a^3),
\end{equation}
where $W=e^{\phi_s}-4e^{-2\phi_s}$. The corresponding bilinear map, evaluated on two arbitrary perturbations $\mathbf a_1=(\eta_1,\pi_1)^{\mathrm T}$ and $\mathbf a_2=(\eta_2,\pi_2)^{\mathrm T}$, is therefore
\begin{equation}\label{ANdefinitions}
	N(\mathbf a_1,\mathbf a_2)
	=
	\begin{pmatrix}
		0\\-W\eta_1\eta_2
	\end{pmatrix}.
\end{equation}
The map $P$ is defined analogously through the quadratic expansion of $Y$. Its explicit expression will not be required.

We now consider the positive-frequency shape mode. Writing $\sigma_1=-i\omega_1=-3i/2$, its time dependence is $e^{\sigma_1t}$. Factoring out this time dependence, the corresponding vector of Cauchy data is
\begin{equation}\label{phasev}
	\mathbf v=
	\begin{pmatrix}
		\chi_1\\
		\sigma_1\chi_1
	\end{pmatrix}.
\end{equation}
As established above, $\mathbf v$ is a common eigenvector of the linearised physical and higher-symmetry generators
\begin{equation}\label{commonv}
	A\mathbf v=\sigma_1\mathbf v,
	\qquad
	B\mathbf v=\mu_1\mathbf v,
	\qquad \mu_1=27i.
\end{equation}

We now exploit the commutativity of the full nonlinear vector fields. As discussed in connection with \eqref{G5symmetry}, the modified characteristic $\mathcal K$ generates a generalised symmetry of the Bullough-Dodd equation. Consequently, its action is compatible with physical time evolution. In terms of the corresponding vector fields $X$ and $Y$ on the space of Cauchy data, this compatibility is expressed by
\begin{equation}\label{XYcommutativity}
	DX(\mathbf a)Y(\mathbf a) -DY(\mathbf a)X(\mathbf a)=0.
\end{equation}
Here $DX(\mathbf a)Y(\mathbf a)$ denotes the infinitesimal variation of $X$ induced by the change $\mathbf a\mapsto \mathbf a+\alpha Y(\mathbf a)$, and similarly for the second term. At linear order, \eqref{XYcommutativity} reproduces $[A,B]=0$, as established above. We now examine the additional constraints arising at quadratic order.

Substituting \eqref{flowexpansions} and collecting the quadratic terms along $\mathbf v$, we obtain
\begin{equation}
		0= \frac12A P(\mathbf v,\mathbf v) +N(\mathbf v,B\mathbf v) -\frac12B N(\mathbf v,\mathbf v) -P(\mathbf v,A\mathbf v).
\end{equation}
Using \eqref{commonv} and the bilinearity of $N$ and $P$ gives
\begin{equation}\label{quadraticcompatibility}
	(B-2\mu_1)N(\mathbf v,\mathbf v)
	=(A-2\sigma_1)P(\mathbf v,\mathbf v).
\end{equation}
The factors of two arise because the quadratic interaction contains two copies of the shape mode. Equation \eqref{quadraticcompatibility} therefore relates the nonlinear sources at twice the physical and higher-symmetry eigenvalues of the leading oscillation.

We now project this identity onto the second-harmonic continuum. At $\omega=2\omega_1=3$, corresponding to $\kappa=\sqrt6$, the dual Jost states constructed above satisfy
\begin{equation}\label{dualresonance}
		A^{\mathrm T}\widetilde{\mathbf v}^{(2)}_\pm =2\sigma_1\widetilde{\mathbf v}^{(2)}_\pm, \qquad B^{\mathrm T}\widetilde{\mathbf v}^{(2)}_\pm =\widetilde\mu_\pm^{(2)} \widetilde{\mathbf v}^{(2)}_\pm.
\end{equation}
Here the pair $\{\widetilde\mu_\pm^{(2)}\}$ coincides with the higher-symmetry eigenvalues in \eqref{musecond}, with the momentum labels interchanged.

Taking the bilinear pairing of \eqref{quadraticcompatibility} with either dual state, the right-hand side vanishes because
\begin{equation}
		\big\langle \widetilde{\mathbf v}^{(2)}_\pm, (A-2\sigma_1)P(\mathbf v,\mathbf v) \big\rangle = \big\langle (A^{\mathrm T}-2\sigma_1) \widetilde{\mathbf v}^{(2)}_\pm, P(\mathbf v,\mathbf v) \big\rangle=0.
\end{equation}
Applying the same transposition to the left-hand side yields
\begin{equation}\label{selectionrule}
	\big(\widetilde\mu_\pm^{(2)}-2\mu_1\big) \big\langle\widetilde{\mathbf v}^{(2)}_\pm, N(\mathbf v,\mathbf v)\big\rangle=0.
\end{equation}

This is the quadratic selection rule. A nonzero coupling to a continuum channel would require its higher-symmetry eigenvalue to coincide with $2\mu_1$. However, as established in \eqref{musecond}, neither continuum eigenvalue satisfies this condition. Consequently,
\begin{equation}\label{selectionzero}
	\big\langle\widetilde{\mathbf v}^{(2)}_\pm, N(\mathbf v,\mathbf v)\big\rangle=0.
\end{equation}
Thus the quadratic shape-mode interaction has vanishing projections onto both on-shell continuum channels.

To make contact with the radiation amplitudes derived previously in  section 3 we recall that $\widetilde{\mathbf v}^{(2)}_\pm =(2\sigma_1q_\pm^{(2)},q_\pm^{(2)})^{\mathrm T}$. Using \eqref{ANdefinitions}, we find
\begin{equation}\label{sourcebridge}
		\big\langle \widetilde{\mathbf v}^{(2)}_\pm, N(\mathbf v,\mathbf v) \big\rangle = -\int_{-\infty}^{\infty} q_\pm^{(2)}W\chi_1^2\,\mathrm dx = 4\int_{-\infty}^{\infty} q_\pm^{(2)}S\,\mathrm dx,
\end{equation}
where $S=-W\chi_1^2/4$ is the
second-harmonic source introduced earlier.

Since the two Jost solutions span the same scattering solution space as $q_e$ and $q_o$, the vanishing projections imply
\begin{equation}
	\mathcal M_e^{(2)}
	=\mathcal M_o^{(2)}=0.
\end{equation}
We have therefore recovered the exact second-harmonic cancellations of section~3 without using the explicit particular solution $\psi_2$. Their origin can be traced to the incompatibility between the higher-symmetry eigenvalue of the quadratic shape-mode interaction and those of the resonant continuum channels.

\subsection{Cubic selection rule and the intermediate response}

We now extend the selection-rule argument to the third harmonic. Unlike the quadratic case, the third-harmonic source contains two contributions: the direct cubic self-interaction of the shape mode and its quadratic interaction with the second-order response. The higher-symmetry constraint must therefore be applied to their sum.

To account for these contributions, we extend the expansions of the commuting vector fields to cubic order
\begin{equation}\label{ flowscubic}
	\begin{aligned}
		X(\mathbf a)
		&=A\mathbf a+\frac12N(\mathbf a,\mathbf a)
		+\frac16R(\mathbf a,\mathbf a,\mathbf a)
		+O(\mathbf a^4),\\
		Y(\mathbf a)
		&=B\mathbf a+\frac12P(\mathbf a,\mathbf a)
		+\frac16Q(\mathbf a,\mathbf a,\mathbf a)
		+O(\mathbf a^4),
	\end{aligned}
\end{equation}
where $R$ and $Q$ are symmetric trilinear maps, defined analogously to the bilinear maps $N$ and $P$ introduced above. For the physical evolution we have
\begin{equation}\label{physicalcubicmap}
	R(\mathbf a_1,\mathbf a_2,\mathbf a_3)
	=
	\begin{pmatrix}
		0\\-Z\eta_1\eta_2\eta_3
	\end{pmatrix},
\end{equation}
with $\mathbf a_j=(\eta_j,\pi_j)^{\mathrm T}$ and $Z=e^{\phi_s}+8e^{-2\phi_s}$. As before, the explicit expressions for the higher-symmetry maps $P$ and $Q$ will not be required.

We first consider the intermediate second-order response. Recall that the leading oscillation was normalised in section~2 as $\varepsilon\chi_1\cos(\omega_1t)$. Setting $h=\varepsilon/2$ and retaining the positive-frequency contributions, the relevant terms in the Cauchy-data expansion take the form
\begin{equation}
	\mathbf a
	=h\mathbf v e^{\sigma_1t}
	+h^2\mathbf w e^{2\sigma_1t}+\cdots.
\end{equation}
Since the positive-frequency part of $\varepsilon^2\psi_2\cos(2\omega_1t)$ is $2h^2\psi_2e^{2\sigma_1t}$, the corresponding second-order coefficient is
\begin{equation}\label{wsecond}
	\mathbf w=
	\begin{pmatrix}
		2\psi_2\\
		4\sigma_1\psi_2
	\end{pmatrix}.
\end{equation}
Substituting this expansion into the physical evolution equation at order $h^2e^{2\sigma_1t}$ gives
\begin{equation}\label{wphysical}
	(A-2\sigma_1)\mathbf w
	=-\frac12N(\mathbf v,\mathbf v),
\end{equation}
which is equivalent to the second-harmonic equation $(H-9)\psi_2=S$ derived earlier.

For the cubic selection rule, we also require the corresponding relation under the higher symmetry,
\begin{equation}\label{whigher}
	(B-2\mu_1)\mathbf w
	=-\frac12P(\mathbf v,\mathbf v).
\end{equation}
Unlike the eigenvalue equations for $\mathbf v$, this relation does not follow directly from linearisation, because $\mathbf w$ is a forced response rather than a linear eigenmode. We can, however, establish it using the quadratic compatibility relation.

To this end, we introduce the defect
\begin{equation}\label{wdefect}
	\mathbf d=(B-2\mu_1)\mathbf w +\frac12P(\mathbf v,\mathbf v).
\end{equation}
Using $[A,B]=0$ and \eqref{wphysical}, we obtain
\begin{equation}
		(A-2\sigma_1)\mathbf d ={}-\frac12(B-2\mu_1) N(\mathbf v,\mathbf v) +\frac12(A-2\sigma_1) P(\mathbf v,\mathbf v).
\end{equation}
The right-hand side vanishes by
\eqref{quadraticcompatibility}, so that
\begin{equation}\label{defecthomogeneous}
	(A-2\sigma_1)\mathbf d=0.
\end{equation}
Because the higher symmetry is a local differential expression, $\mathbf d$ is spatially localised whenever $\mathbf v$ and $\mathbf w$ are. Writing $\mathbf d=(d_1,d_2)^{\mathrm T}$, the homogeneous equation implies $(H-9)d_1=0$. At this positive continuum energy, the fluctuation operator has no nontrivial spatially localised homogeneous solution. Hence $\mathbf d=0$, establishing \eqref{whigher}.

We have therefore shown that the localised second-order response is compatible with both commuting evolutions. This property allows us to incorporate its contribution into the cubic selection rule.

We now expand the commutativity condition \eqref{XYcommutativity} to cubic order. Evaluating the resulting identity on three copies of $\mathbf v$ gives
\begin{equation}\label{cubiccommutator}
		(B-3\mu_1)R(\mathbf v,\mathbf v,\mathbf v) -(A-3\sigma_1)Q(\mathbf v,\mathbf v,\mathbf v)
		=3\left[ N\big(\mathbf v,P(\mathbf v,\mathbf v)\big) -P\big(\mathbf v,N(\mathbf v,\mathbf v)\big) \right].
\end{equation}
This identity constrains the direct cubic interaction, but does not yet incorporate the intermediate second-order response.

To include this contribution, we polarise the quadratic part of the commutativity condition, taking $\mathbf v$ and $\mathbf w$ as its two arguments. Using the eigenvalue relations \eqref{commonv} and the response equations \eqref{wphysical} and \eqref{whigher}, we obtain
\begin{equation}\label{polarisedcompatibility}
		(B-3\mu_1)N(\mathbf v,\mathbf w) -(A-3\sigma_1)P(\mathbf v,\mathbf w)
		= -\frac12N\big(\mathbf v,P(\mathbf v,\mathbf v)\big) +\frac12P\big(\mathbf v,N(\mathbf v,\mathbf v)\big).
\end{equation}
Adding one sixth of \eqref{cubiccommutator} to \eqref{polarisedcompatibility}, the remaining terms cancel. We arrive at the cubic compatibility relation
\begin{equation}\label{cubicmaster}
	(B-3\mu_1)\mathbf F_X^{(3)}=(A-3\sigma_1)\mathbf F_Y^{(3)},
\end{equation}
where
\begin{equation}\label{cubicsources}
		\mathbf F_X^{(3)}
		=N(\mathbf v,\mathbf w)
		+\frac16R(\mathbf v,\mathbf v,\mathbf v), \qquad
		\mathbf F_Y^{(3)}
		=P(\mathbf v,\mathbf w)
		+\frac16Q(\mathbf v,\mathbf v,\mathbf v).
\end{equation}
This relation is the cubic counterpart of \eqref{quadraticcompatibility}. Importantly, it constrains the complete third-harmonic forcing, including both the direct cubic interaction and the intermediate second-order response.

Using \eqref{ANdefinitions}, \eqref{physicalcubicmap} and the normalisation \eqref{wsecond}, the physical source becomes
\begin{equation}\label{cubicphysicalsource}
	\begin{aligned}
		\mathbf F_X^{(3)}
		&=
		\begin{pmatrix}
			0\\
			-2W\chi_1\psi_2-\frac16Z\chi_1^3
		\end{pmatrix}
		=\begin{pmatrix}0\\4S_3(x)\end{pmatrix},\\
		S_3&=-\frac12W\chi_1\psi_2
		-\frac1{24}Z\chi_1^3.
	\end{aligned}
\end{equation}
This agrees with the third-harmonic source introduced in section~3. The Poincar\'e-Lindstedt frequency correction contributes only to the fundamental harmonic at this order and therefore does not enter $S_3$.

The final step follows the same projection argument as in the quadratic case. At the third-harmonic frequency $\omega=3\omega_1=9/2$, the continuum wavenumber is $\kappa_3=\sqrt{69}/2$. The corresponding dual Jost states satisfy
\begin{equation}\label{dualmodethird}
		A^{\mathrm T}\widetilde{\mathbf v}^{(3)}_\pm
		=3\sigma_1\widetilde{\mathbf v}^{(3)}_\pm, \qquad
		B^{\mathrm T}\widetilde{\mathbf v}^{(3)}_\pm
		=\widetilde\mu_\pm^{(3)}
		\widetilde{\mathbf v}^{(3)}_\pm,
\end{equation}
where the pair $\{\widetilde\mu_\pm^{(3)}\}$ coincides with the eigenvalues in \eqref{muthird}.

Taking the bilinear pairing of \eqref{cubicmaster} with either dual state and using these eigenvalue equations, we obtain
\begin{equation}\label{cubicselection}
	\big(\widetilde\mu_\pm^{(3)}-3\mu_1\big)
	\big\langle\widetilde{\mathbf v}^{(3)}_\pm,
	\mathbf F_X^{(3)}\big\rangle=0.
\end{equation}
Since $\widetilde\mu_\pm^{(3)}\ne3\mu_1$, both projections must vanish.

Using the explicit dual vectors $\widetilde{\mathbf v}^{(3)}_\pm =(3\sigma_1q_\pm^{(3)},q_\pm^{(3)})^{\mathrm T}$ and \eqref{cubicphysicalsource}, this condition becomes
\begin{equation}\label{cubicprojectionbridge}
	\big\langle\widetilde{\mathbf v}^{(3)}_\pm,
	\mathbf F_X^{(3)}\big\rangle
	=4\int_{-\infty}^{\infty}
	q_\pm^{(3)}(x)S_3(x)\,\dd x=0.
\end{equation}
Since the two Jost solutions span the same scattering solution space as $q_e$ and $q_o$, we conclude that
\begin{equation}\label{thirdhighersymmetryzero}
	\mathcal M_e^{(3)}=\mathcal M_o^{(3)}=0.
\end{equation}

The fifth-order generalised symmetry thus enforces the cancellation of the complete third-harmonic radiation source. The argument requires the localised second-order response $\psi_2$, but not the explicit third-harmonic particular solution $\psi_3$. Together with the quadratic selection rule, this provides a symmetry-based explanation for the two on-shell cancellations established directly in section~3 using Lagrange's identity.

\section{Conclusions}

We have investigated nonlinear excitations of the discrete shape mode about static complex type-I Bullough-Dodd solitons. Extending the linear fluctuation analysis to third order in perturbation theory, we have constructed explicit spatially localised corrections at the static, fundamental, second-harmonic and third-harmonic frequencies. The Poincar\'e-Lindstedt method removes the secular contribution at the fundamental frequency and determines the leading nonlinear frequency shift, $\Omega=\omega_1+9\varepsilon^2/32+O(\varepsilon^3)$. Remarkably, the equations governing both higher harmonics admit exact spatially localised particular solutions.

The significance of these solutions becomes apparent from the fluctuation spectrum. Both the second and third harmonics lie strictly inside the continuous spectrum, where propagating radiation channels are available. Nevertheless, the nonlinear sources at these frequencies have exactly vanishing on-shell overlaps with both independent scattering channels. We have established these cancellations directly using Lagrange's identity, which relates the vanishing radiation amplitudes to the existence of the localised particular solutions. Thus, the absence of radiation at these perturbative orders is not due to the absence of kinematically accessible continuum modes, but to the vanishing of their nonlinear excitation amplitudes.

To understand the origin of these cancellations, we have examined the fifth-order generalised symmetry of the integrable Bullough-Dodd equation. By combining this symmetry with a translation, we constructed a modified higher symmetry that leaves the static soliton invariant. Its linearisation provides an additional eigenvalue characterising the shape mode and the continuum scattering states. The central observation is that resonance under physical time evolution does not imply resonance under this higher symmetry. At the second and third harmonics, the higher-symmetry eigenvalues of the continuum channels do not match those associated with the corresponding nonlinear shape-mode interactions.

The commutativity of the physical and higher-symmetry evolutions translates this eigenvalue mismatch into selection rules for the nonlinear source projections. At quadratic order, the constraint directly eliminates the on-shell coupling to the second-harmonic continuum. At cubic order, the analysis must also incorporate the localised second-order response, whose interaction with the leading shape mode contributes to the third-harmonic forcing. The resulting compatibility relation constrains the complete cubic source and again forces both on-shell radiation amplitudes to vanish. The higher-symmetry analysis therefore provides a structural interpretation of the cancellations in terms of integrability, beyond their direct verification using explicit solutions.

This behaviour contrasts with the familiar nonintegrable $\phi^4$ theory, where excitation of the kink's internal mode produces second-harmonic radiation and consequently radiative damping \cite{manton1997kinks,barashenkov2009wobbling}. In the present model, the absence of radiation at the corresponding perturbative orders is not explained by the fluctuation spectrum, since the relevant continuum channels are kinematically accessible. Nor is it an automatic consequence of integrability. Rather, the cancellations follow from specific constraints imposed by the fifth-order commuting symmetry. In particular, the cubic selection rule also accounts for the intermediate second-order response, showing that the complete third-harmonic source is subject to the same symmetry constraint. This provides a concrete mechanism through which integrability can suppress nonlinear radiation that would otherwise be permitted.

These results are restricted to the perturbative orders considered here. In particular, they do not establish full nonlinear stability or the existence of an exact spatially localised, time-periodic excitation. Radiation from more general initial conditions, contributions at higher perturbative orders and limitations on the long-time validity of the expansion remain possible. Furthermore, while the higher-symmetry argument is supported by the scattering properties of the linearised fluctuation operator, a fully rigorous functional-analytic formulation would require a systematic treatment of the dual continuum eigenchannels and their spectral projections.

An important question for future work is whether the observed cancellations persist at all perturbative orders. If so, it would be natural to investigate whether the resulting expansion corresponds to an exact family of spatially localised, time-periodic solutions. More generally, our results illustrate how higher integrable symmetries can impose constraints on nonlinear mode coupling that are not apparent from the physical fluctuation spectrum alone. Understanding the extent of these constraints may provide further insight into the suppression of radiation in integrable field theories.

\section*{Appendix}
\appendix

\section{Higher conserved density and its relation to the fifth-order symmetry}
\label{app:BDcharge}

The selection rules derived in section~4 rely on the fifth-order generalised symmetry of the Bullough-Dodd equation. For completeness, we show here that its characteristic is also directly related to a higher conserved density of the integrable hierarchy. We first verify the corresponding conservation law and then establish the algebraic relation between the density and the symmetry characteristic.

We use the light-cone notation introduced in \eqref{lightcone} and \eqref{BDlightcone}. Recalling $u_n=D_+^n\phi$, commutativity of the light-cone derivatives gives, on solutions of the field equation,
\begin{equation}\label{jetrelation}
	D_-u_n=D_+^{n-1}f(\phi),
	\qquad n\geq1.
\end{equation}
In particular,
\begin{equation}\label{jetexamples}
		D_-u_1=f, \qquad 
		D_-u_2=f'u_1,  \qquad
		D_-u_3=f''u_1^2+f'u_2.
\end{equation}
We will also use $f''''=5f''-4f$, which follows directly from (\ref{fidentity}).

Next we consider the local expressions
\begin{equation}\label{chiralchargeexpressions}
	\begin{aligned}
		T_6={}&u_3^2-\frac53u_2^3 +5u_1^2u_2^2+\frac13u_1^6,\\
		J_4={}&2f''u_1^2u_2+f'u_2^2 +\frac12(5f'-f''')u_1^4.
	\end{aligned}
\end{equation}
We shall verify that these satisfy a local conservation law. Applying $D_-$ to $T_6$, and using \eqref{jetexamples}, gives
\begin{equation}\label{DminusT6}
		D_-T_6=(10f-5f')u_1u_2^2 +10f'u_1^3u_2+2fu_1^5 +2f''u_1^2u_3+2f'u_2u_3.
\end{equation}
On the other hand, applying $D_+$ to the flux $J_4$ yields
\begin{equation}\label{DplusJ4}
		D_+J_4=5f''u_1u_2^2 +10f'u_1^3u_2 +\frac12(5f''-f'''')u_1^5 +2f''u_1^2u_3+2f'u_2u_3.
\end{equation}
The two expressions coincide when using the above expression for $f''''$. We therefore obtain
the local conservation law
\begin{equation}\label{chiralconservation}
	D_-T_6=D_+J_4.
\end{equation}

Returning to the original space and time coordinates, this identity becomes
\begin{equation}\label{chargecontinuity}
	\partial_t(T_6-J_4)
	=\partial_x(T_6+J_4).
\end{equation}
Consequently, the quantity
\begin{equation}\label{highercharge}
	Q_5=\int_{-\infty}^{\infty}
	(T_6-J_4)\,\mathrm dx
\end{equation}
is conserved, provided the boundary contribution $[T_6+J_4]_{-\infty}^{\infty}$ vanishes. In particular, this condition is satisfied for sufficiently smooth field configurations whose relevant derivatives decay at spatial infinity and whose asymptotic values belong to the vacuum set of the theory. The vacuum values at the two spatial infinities need not coincide.

The conserved density $T_6$ is directly related to the fifth-order generalised symmetry \eqref{G5higher}. Treating $u_1$ as the dependent variable, its formal Euler derivative is
\begin{equation}
		\frac{\delta T_6}{\delta u_1} = \frac{\partial T_6}{\partial u_1} -D_+\frac{\partial T_6}{\partial u_2} +D_+^2\frac{\partial T_6}{\partial u_3} =2\mathcal G_5[\phi].
\end{equation}
This identity establishes an algebraic connection between the higher conserved density and the generalised symmetry underlying the selection rules of section~4. It does not, however, imply that the conserved charge $Q_5$ generates this symmetry through the physical Poisson bracket, which would require a separate Hamiltonian analysis.


\begin{thebibliography}{99}
	
\bibitem{manton1997kinks}
	N.~Manton and H.~Merabet,
	\newblock $\phi^4$ kinks: Gradient flow and dynamics,
	\newblock Nonlinearity {\bf 10}, 3--18 (1997).
	
\bibitem{barashenkov2009wobbling}
	I.~Barashenkov and O.~Oxtoby,
	\newblock Wobbling kinks in $\phi^4$ theory,
	\newblock Phys. Rev. E {\bf 80}, 026608 (2009).
	
\bibitem{takyi2016coll}
	I.~Takyi and H.~Weigel,
	\newblock Collective coordinates in one-dimensional soliton models revisited,
	\newblock Phys. Rev. D {\bf 94}, 085008 (2016).
	
\bibitem{navarro2023inclusion}
	S.~Navarro-Obreg{\'o}n, L.~Nieto and J.~Queiruga,
	\newblock Inclusion of radiation in the collective coordinate method
	approach of the $\phi^4$ model,
	\newblock Phys. Rev. E {\bf 108}, 044216 (2023).
	
\bibitem{kowalczyk2017kink}
	M.~Kowalczyk, Y.~Martel and C.~Mu{\~n}oz,
	\newblock Kink dynamics in the $\phi^4$ model:
	Asymptotic stability for odd perturbations in the energy space,
	\newblock J. Am. Math. Soc. {\bf 30}, 769--798 (2017).
	
\bibitem{KowalczykMartel2024}
	M.~Kowalczyk and Y.~Martel,
	\newblock Kink dynamics under odd perturbations for (1+1)-scalar
	field models with one internal mode,
	\newblock Math. Res. Lett. {\bf 31}, 795--832 (2024).
	
\bibitem{soffer1999resonances}
	A.~Soffer and M.~I. Weinstein,
	\newblock Resonances, radiation damping and instability in
	Hamiltonian nonlinear wave equations,
	\newblock Invent. Math. {\bf 136}, 9--74 (1999).
	
\bibitem{Parke1980}
	S.~Parke,
	\newblock Absence of particle production and factorization
	of the S-matrix in 1+1 dimensional models,
	\newblock Nucl. Phys. B {\bf 174}, 166--182 (1980).
	
\bibitem{dorey2022tree}
	P.~Dorey and D.~Polvara,
	\newblock Tree-level integrability in 2d quantum field theories
	and affine Toda models,
	\newblock JHEP {\bf 02} (2022) 199.
	
\bibitem{AKNSsg}
	M.~J. Ablowitz, D.~J. Kaup, A.~C. Newell and H.~Segur,
	\newblock Method for solving the sine-Gordon equation,
	\newblock Phys. Rev. Lett. {\bf 30}, 1262 (1973).
	
\bibitem{CorreaFringTaira}
	F.~Correa, A.~Fring and T.~Taira,
	\newblock Linearly stable and unstable complex soliton solutions
	with real energies in the Bullough-Dodd model,
	\newblock Nucl. Phys. B {\bf 979}, 115783 (2022).
	
\bibitem{tzitzeica1908}
	G.~Tzitz{\'e}ica,
	\newblock Sur une nouvelle classe de surfaces,
	\newblock Rend. Circ. Mat. Palermo {\bf 25}, 180--187 (1908).
	
\bibitem{BDodd}
	R.~Dodd and R.~Bullough,
	\newblock Polynomial conserved densities for the sine-Gordon equations,
	\newblock Proc. R. Soc. Lond. A {\bf 352}, 481--503 (1977).
	
\bibitem{zhiber1979klein}
	A.~V. Zhiber and A.~B. Shabat,
	\newblock The Klein--Gordon equation with nontrivial group,
	\newblock Dokl. Akad. Nauk SSSR {\bf 247}, 1103--1107 (1979).
	
\bibitem{DIO1983}
	D.~I. Olive and N.~Turok,
	\newblock The symmetries of Dynkin diagrams and the reduction
	of Toda field equations,
	\newblock Nucl. Phys. B {\bf 215}, 470 (1983).
	
\bibitem{andreevback}
	V.~A. Andreev,
	\newblock B{\"a}cklund transformations of the
	Bullough-Dodd-Zhiber-Shabat equation and symmetries
	of integrable equations,
	\newblock Theor. Math. Phys. {\bf 79}, 448--450 (1989).
	
\bibitem{assis2008bullough}
	P.~E.~G. Assis and L.~A. Ferreira,
	\newblock The Bullough--Dodd model coupled to matter fields,
	\newblock Nucl. Phys. B {\bf 800}, 409--449 (2008).
	
\bibitem{Fring:1992pj}
	A.~Fring, G.~Mussardo and P.~Simonetti,
	\newblock Form factors of the elementary field in the
	Bullough-Dodd model,
	\newblock Phys. Lett. B {\bf 307}, 83--90 (1993).
	
\bibitem{Amila}
	A.~Poric,
	\newblock Stability of soliton solutions,
	\newblock MSc dissertation, City St George's, University of London (2026).
	
\bibitem{boyce2026elementary}
	W.~E. Boyce, R.~C. DiPrima and D.~B. Meade,
	\newblock {\em Elementary Differential Equations and
		Boundary Value Problems},
	\newblock 12th edition, Wiley (2021).
	
\bibitem{vNW}
	J.~von Neumann and E.~Wigner,
	\newblock {\"U}ber merkw{\"u}rdige diskrete Eigenwerte,
	\newblock Phys. Z. {\bf 30}, 465--467 (1929).
	
\bibitem{bender2013advanced}
	C.~M. Bender and S.~A. Orszag,
	\newblock {\em Advanced Mathematical Methods for Scientists
		and Engineers I: Asymptotic Methods and Perturbation Theory},
	\newblock Springer (2013).
	
\bibitem{kaptsov2018int}
	O.~V. Kaptsov,
	\newblock Intermediate systems and higher-order
	differential constraints,
	\newblock J. Sib. Fed. Univ. Math. Phys. {\bf 11},
	550--560 (2018).
	
\end{thebibliography}
\end{document}